\documentclass[twocolumn,amsmath,amssymb,nofootinbib]{revtex4}
\usepackage{graphicx}
\usepackage{dcolumn}
\usepackage{color}
\usepackage{bm}
\usepackage{amsmath,amssymb,graphicx,hyperref,xcolor}
\usepackage{epsfig}
\usepackage{enumerate}
\usepackage{array}
\usepackage{multirow}

\begin{document}
\title
{Landauer's principle and black hole area quantization}
\author{Bijan Bagchi\footnote{E-mail: bbagchi123@gmail.com}$^1$,  
Aritra Ghosh\footnote{E-mail: ag34@iitbbs.ac.in}$^2$, 
Sauvik Sen\footnote{E-mail: sauviksen.physics@gmail.com}$^3$}

\vspace{2mm}

\affiliation{$^{1}$Department of Mathematics, Brainware University,
Barasat, Kolkata, West Bengal 700125, India\\
$^{2}$School of Basic Sciences, Indian Institute of Technology Bhubaneswar, Jatni, Khurda, Odisha 752050, India\\
$^3$ Department of Physics, Shiv Nadar Institution of Eminence, Gautam
Buddha Nagar, Uttar Pradesh 203207, India}

\vskip-2.8cm
\date{\today}
\vskip-0.9cm


\begin{abstract}
This article assesses Landauer's principle from information theory in the context of area quantization of the Schwarzschild black hole. Within a quantum-mechanical perspective where Hawking evaporation can be interpreted in terms of transitions between the discrete states of the area (or mass) spectrum, we justify that Landauer's principle holds consistently in the saturated form when the number of microstates of the black hole goes as \(2^n\), where \(n\) is a large positive integer labeling the levels of the area/mass spectrum in the semiclassical regime. This is equivalent to the area spacing \(\Delta A = \alpha l_P^2\) (in natural units), where \(\alpha = 4 \ln 2\) for which the entropy spacing between consecutive levels in Boltzmann units coincides exactly with one bit of information. We also comment on the situation for other values of \(\alpha\) prevalent in the literature.
\end{abstract}

\maketitle

\textbf{Introduction:} In 1961, while dealing with dissipation and minimal heat generation, Landauer \cite{land1} provided an elegant thermodynamical reasoning to interpret digital systems to the effect that removal of one bit of information is needed to be accompanied by a minimum amount of energy loss which is proportional to the governing temperature of the operating system. Presupposing that information is physical and that it possesses an energy equivalence \cite{land2, land3, borma1}, erasure of information could be interpreted as being kind of a dissipative phenomenon \cite{berut}. This idea has been generalized to many-valued logics (see for example, \cite{borma}), while more recently, relativistic communication possibilities have been explored \cite{alvim}. 
 Landauer's paper has received widespread attention (see \cite{benn} for an excellent review which also introduces many other useful facets to the principle; also see \cite{raju} for a description of the information paradox related to black hole thermodynamics) in the context of a computer being regarded as a closed system admitting a Hamiltonian formulation which is controlled by unitary dynamics. The key point made by Landauer was that the evolution of a logical state is likely to be irreversible with multiple logical states having a single logical
successor. The connection between logical irreversibility
and entropy changes points towards the fact that a decrease in entropy (pertaining to information-bearing degrees of freedom) should signal a corresponding compensation through the increase of entropy in the non-information-bearing degrees of freedom. However, it may also be noted that the authors of \cite{Ancilla} have suggested that Ancilla-assisted erasure of quantum information may violate the Landauer bound.  \\

In a recent enlightening work on Hawking radiation, Cort\^{e}s and Liddle \cite{liddle} have re-examined Landauer's principle of information thermodynamics to identify the temperature with the Hawking temperature $T_{\rm H}$ of a Schwarzschild black hole and have observed that the energy dissipated to the surroundings coincides exactly with the Landauer energy for radiation at the black hole temperature upon using the \(\ln 2\) factor to convert between entropy in Boltzmann units and information in bits. This result has been termed by them as the `black hole saturation of the Landauer bound'. A follow-up study on the correspondence between information dynamics and expanding cosmic horizons has also been reported \cite{trivedi}.\\

Let us recall that following Bekenstein's conjecture \cite{bek1, bek2} that the area $A$ of the event horizon could be quantized as being proportional to the square of the Planck length $l_P$, a precise estimate of it was given as $A = 16 \pi \left (\frac{GM}{c^2} \right )^2$  \cite{bek4,bek3}, where \(G\) is the Newton's constant, \(M\) is the mass of the black hole, and \(c\) is the speed of light in vacuum. Such a formula on account of adiabatic invariants implies a uniform spacing of $A$ which is in tune with the Bohr-Sommerfeld
quantization rule and which in turn gives discrete values for $M$. The natural expectation is then that as the black hole evaporates, transitions could occur between discrete values of mass states leading to emissions of radiation \cite{lou}. The idea of a black hole radiating thermally \cite{haw1, haw2} means that it could decrease in mass \cite{silk}, and as such one could assign an entropy $S_{\rm BH}$ which is related to $A$ in the following manner:
\begin{equation}\label{2}
    \frac{S_{\rm BH}}{k_B} = \frac{c^3}{4 \hbar G} A = \frac{4 \pi G}{c\hbar} M^2,
\end{equation}
where \(k_B\) is the Boltzmann constant and \(\hbar\) is the (reduced) Planck's constant. It clearly reveals that the horizon area of the black hole also increases irreversibly when matter is absorbed with $A$ serving as a measure of information \cite{bek2}. The corresponding temperature, namely, the Hawking temperature, can be easily recognized to be
\begin{equation}\label{3}
    k_B T_{\rm H} = \frac{\hbar c^3}{8 \pi G M} ,
\end{equation}
pointing to $S_{\rm BH}$ and $T_{\rm H}$ both being `quantum mechanical' in character owing to the explicit presence of $\hbar$. However, the product $S_{\rm BH} T_{\rm H}$ is independent of $\hbar$. We wish to comment here that in the derivation of Eqs. (\ref{2}) and (\ref{3}), gravity was accounted for classically while non-gravitational fields were considered from a quantum perspective\footnote{More precisely, Hawking considered a classical (background) spacetime for the gravitational collapse to a Schwarzschild black hole with a free quantum field propagating in it \cite{wald}.}. Such an approximation of course breaks down when the dimension of the black hole shrinks to the Planck scale\footnote{The Planck length goes as $l_P = \sqrt{\frac{G\hbar}{c^3}}$.} where the rules of quantum gravity apply \cite{kiefer}. The study of black hole entropy for this microscopic regime would give a truer picture of the statistical mechanics of gravitational degrees of freedom. A microscopic interpretation of black hole entropy focuses on the von Neumann
formula; Bekenstein and
Mukhanov \cite{bek4} provided a quantized treatment of the area of a black hole based on this. \\

The purpose of the present study is to have a look at Landauer's principle in the light of area quantization of black holes which could reveal insights into their quantum nature. Following \cite{liddle}, we shall consider Hawking evaporation of a Schwarzschild black hole but in the spirit of area quantization which has been studied by several researchers earlier \cite{bek4,kiefer,bek5,mukh,quantBH,das2,quant111,quant1111,quant112,quant11,deeg,quant12,quant14,jal} (see also, the papers \cite{rghosh3,kast1,mann,furs,gango,pani}). Here, the evaporation of the black hole occurs in discrete lumps of radiation being emitted due to transitions between the area/mass states. \\

\textbf{Area quantization and Landauer's formula:} Let us consider the semiclassical approximation in which it is now well known that the area spectrum is expressible as (in units where \(\hbar = c = k_B = 1\) and \(G = l_P^2\)) \(\Delta A = \alpha l_P^2\), or
\begin{equation}\label{areaspectrum}
  A \equiv A(n) = \frac{\alpha n}{ m_P^2},
    \end{equation} where \(n\) is a positive integer (\(n >> 1\) for the semiclassical limit), \(m_P = l_P^{-1}\) is the Planck mass, and \(\alpha\) is a suitable constant \cite{bek4,mukh}. This implies that the semiclassical Bekenstein-Hawking entropy turns out to be \(S_{\rm BH} = \alpha n/4\) and further, if it admits a statistical interpretation (as it should), the number of microstates can be expressed as \(\Omega(n) \approx e^{\alpha n/4}\). Since the number of microstates should be a positive integer, we must pick \(\alpha = 4 \ln k\), where \(k = 2,3,4,\cdots\) \cite{bek4}. This essentially gives
    \begin{equation}
    \Omega(n) = k^n. 
    \end{equation} 
    To now obtain Landauer's principle, let us consider the following equality:
    \begin{equation}
    2 \frac{\Delta M}{M} = \frac{\Delta S_{\rm BH}}{S_{\rm BH}}.
    \end{equation}
    Noting that \(M = 2 T_{\rm H} S_{\rm BH}\) for the Schwarzschild black hole, we must have 
     \begin{equation}\label{123456}
    \Delta M = T_{\rm H} \Delta S_{\rm BH},
    \end{equation} which is just the first law of thermodynamics. Thus, the area spectrum implies that
    \begin{equation}
    \Delta M = T_{\rm H} \frac{\alpha}{4} = T_{\rm H} \ln k. 
    \end{equation}
    In the natural units as employed here, the energy `dissipation' accompanying this transition from the state \(|n+1\rangle\) to \(|n\rangle\) is just \(\Delta E = \Delta M\). If \(k = 2\), then the result saturates the Landauer bound \cite{liddle}, i.e., \(\Delta E = T_{\rm H} \ln 2\); here one unit of Bekenstein-Hawking entropy in Boltzmann units is associated with the information content of one bit and this corresponds to \(\Omega(n) \approx 2^n\) for an integer \(n >> 1\), consistent with \cite{bek4,mukh}. \\
    
However, if \(k = 3\) as considered in \cite{quantBH} (see also, \cite{quant111,quant112}), we get \(\Delta E = T_{\rm H} \ln 3\) and in this case, one unit of Bekenstein-Hawking entropy in Boltzmann units does not correspond to an information content of one bit. Here it must also be mentioned that in some works (see for example, \cite{quant12,quant14,jal}), one finds \(\alpha = 8 \pi\) which indicates that \(\Omega(n)\) is not an integer although since one is considering the semiclassical regime for which \(n\) is large, it is difficult to estimate \(\Omega(n)\) with a precision
of order one as is necessary to distinguish an integer from a
non-integer value \cite{quant12}. In this case, one finds \(\Delta E = 2 \pi T_{\rm H}\), which obviously differs from the previously-quoted results. Therefore, we can quote the following result -- \textit{For a Schwarzschild black hole whose area spectrum is given by Eq. (\ref{areaspectrum}), the energy change associated with a transition between consecutive levels is \(\Delta E = T_{\rm H} \alpha/4\), with Landauer's principle emerging in the saturated form for \(\alpha = 4 \ln 2\)}. \\

 \textit{Note --} It may be emphasized that when \(\alpha \neq 4 \ln 2\), one quantum of the black hole entropy does not correspond to one bit of information. While one can still find the associated loss of energy as discussed above, the result does not correspond to Landauer's principle. Since the area levels are quantized, information loss may happen in units other than bits.\\

\textbf{Quantum corrections to temperature:} The spectrum for the black hole area reveals that the mass of the black hole is associated with the following spectrum:
\begin{equation}
M \equiv M(n) = \frac{m_P}{4} \sqrt{\frac{\alpha n}{\pi}}, 
\end{equation}
and consequently, for the transition \(|n+1\rangle \rightarrow |n \rangle\), the change in the black hole mass goes as
\begin{eqnarray}
\Delta M &=& \frac{m_P}{4}\sqrt{\frac{\alpha}{\pi}} \big( \sqrt{n+1} - \sqrt{n} \big) \nonumber \\
&=& \frac{m_P}{4}\sqrt{\frac{\alpha n}{\pi}}\bigg( \sqrt{1 + \frac{1}{n}} - 1 \bigg) \nonumber \\
&\approx&  \frac{m_P}{8}\sqrt{\frac{\alpha}{n \pi}} + \mathcal{O}(m_P n^{-3/2}), \label{DeltaM1111}
\end{eqnarray} where in the last step, we have performed a binomial expansion for \(n >> 1\). This is consistent with Eq. (\ref{123456}). \\

Since the spectrum of the black hole area mimics that of a harmonic oscillator, it may be worthwhile to investigate the effect of a possible zero-point contribution to the area (and mass) spectrum, i.e., we can consider
\begin{equation}
A(n) =  \frac{\alpha}{ m_P^2} \bigg(n + \frac{1}{2}\bigg), \quad M(n) = \frac{m_P}{4} \sqrt{\frac{\alpha}{\pi}\bigg(n + \frac{1}{2}\bigg)},
\end{equation} where it should be remembered that \(n\) is still a large-enough (positive) integer so that the semiclassical results can be trusted. This consistently gives Eq. (\ref{DeltaM1111}) but further, one has
\begin{equation}
n = \frac{16 \pi M^2}{\alpha m_P^2} - \frac{1}{2}, 
\end{equation} which can be combined with Eq. (\ref{DeltaM1111}) to obtain 
\begin{equation}\label{DeltaMcorrected}
\Delta M \approx \frac{m_P}{8}\sqrt{\frac{\alpha}{n \pi}} \approx \frac{\alpha m_P^2}{32 \pi M} \bigg(1 + \frac{\alpha m_P^2}{64 \pi M^2}\bigg). 
\end{equation}
Similarly, one has \(\Delta S_{\rm BH} = \alpha/4\). Then, the first law of thermodynamics leads to a corrected temperature \(\mathcal{T}\) that appears as \(\Delta M = \mathcal{T} \Delta S_{\rm BH}\) and assumes the following expression:
\begin{equation}\label{corrtemp}
\mathcal{T}= \frac{m_P^2}{8 \pi M} \bigg(1 + \frac{\alpha m_P^2}{64 \pi M^2}\bigg) = T_{\rm H} \bigg(1 + \frac{\alpha m_P^2}{64 \pi M^2}\bigg),
\end{equation} where \(T_{\rm H} = \frac{m_P^2}{8 \pi M}\) is the standard expression for the Hawking temperature. From Eq. (\ref{DeltaMcorrected}), we now find
\begin{equation}\label{16}
    \Delta {E} = \mathcal{T} \frac{\alpha}{4} =  T_{\rm H} \frac{\alpha}{4} \bigg(1 + \frac{\alpha m_P^2}{64 \pi M^2}\bigg),
\end{equation}
which obviously satisfies the Landauer's principle for the corrected temperature \(\mathcal{T}\) if \(\alpha = 4 \ln 2\). \\

\textbf{Discussion:} Landauer’s principle is well known to provide deep insights into our understanding of irreversibility and heat generation in computing processes while shedding light on the underpinnings of information erasure. Here we have looked into its relevance with the quantized aspects of a black hole. A few remarks are in order. In the framework of general relativity, the energy content of the universe is taken to be smooth down to the minutest scales of spacetime, although in models of quantum gravity, by implementing modified forms of the uncertainty principle, a minimal-length criterion is supposed to account for high-energy modifications \cite{bagchi1, bosso}. As the black hole diminishes in size and ultimately reaches the Planck size, the consequent remnant would tend to hold an arbitrarily-large amount of information signifying that $S_{\rm BH}$ may cease to be a true signature of the measure of the information content of the black hole. Predictably, the quantum character of a black hole would show similar features as one expects from excited atoms \cite{kot,silk}. Note that a discrete spectrum can have observable effects on the gravitational-wave signals emanating from an inspiraling, binary black hole \cite{datta}. For a different opinion held in the context of gravitational-wave echoes during black hole mergers, see \cite{coates}. \\

Let us also present the possible correction to the black hole entropy due to the modified
temperature defined in Eq. (\ref{corrtemp}). Although the Bekenstein-Hawking entropy still reads \(S_{\rm BH} = A/4l_P^2\), we can define a `corrected' form of the entropy as (see \cite{jal,bagchi2,majhi2} for more details)
\begin{equation}
\mathcal{S} = \int \frac{dM}{\mathcal{T}} = S_{\rm BH} - \frac{\alpha}{16} \ln S_{\rm BH} + \cdots, 
\end{equation}
which exhibits logarithmic corrections. Thus, the coefficient characterizing the logarithmic corrections depends on the parameter \(\alpha\) that characterizes the area spectrum of the black hole. These corrections are of the form \(\mathcal{S} = S_{\rm BH} - c_0 \ln S_{\rm BH} + \cdots\), with (a) \(c_0 = \pi/2\) if \(\alpha = 8\pi\) as in \cite{quant12,quant14,jal,bagchi2}, or (b) \(c_0 = (1/4) \ln k\) if \(\alpha = 4 \ln k\) with \(k = 2, 3, \cdots\) \cite{bek4,mukh,quantBH,quant111,quant112}. We refer the reader to \cite{mann,majhi2,maj1,mitra,sen,partha,ghoshlog} for detailed discussions on logarithmic corrections to the black hole entropy. It should be pointed out that the above-mentioned corrections for \(\alpha = 8\pi\) differ from those presented in \cite{jal,bagchi2} by a factor of 1/2; the latter corrections were obtained based on the Stefan-Boltzmann law.\\

 Finally, let us discuss the effect of the inclusion of angular momentum as in the Kerr black hole whose mass reads as (we take \(G = l_P^2 = 1\) so that our equations are consistent with those of \cite{126})
\begin{equation}
M = \sqrt{\frac{A}{16 \pi} + \frac{4 \pi J^2}{A} }.
\end{equation} 
Here, in addition to quantized area levels as suggested in Eq. (\ref{areaspectrum}), we would also encounter the quantization of angular momentum, i.e., \(J = j\), where \(j\) is a semi-integer bounded by \(0\leq j \leq \alpha n/8 \pi\). Consequently, the mass spectrum turns out to be 
\begin{equation}
M(n,j) = \sqrt{ \frac{\alpha n}{16 \pi} + \frac{4 \pi j^2}{\alpha n}}.
\end{equation}
For the sake of concreteness, let us consider the situation taken up in \cite{126} where transitions conforming to \(\Delta j = 2\) were focused on from the point of view of the dominant gravitational-wave mode which is the most interesting for astrophysical systems. In that case, the transition between \(M(n+1,j+2)\) and \(M(n,j)\) yields the change in energy quantified by (see also, \cite{datta,126,PRD2})
\begin{equation}
\Delta E = T_{\rm H} \frac{\alpha }{4} + 2 \Omega_{\rm H},
\end{equation} where \(\Omega_{\rm H}\) is the angular velocity and we have taken \(n >> 1\). For the Kerr black hole, we have the following well-known expressions:
\begin{equation}\label{rotatingtempexp}
T_{\rm H} = \frac{\sqrt{1 - a^2}}{4 \pi M (1 + \sqrt{1 - a^2})}, \quad \quad \Omega_{\rm H} = \frac{a}{2 M (1 + \sqrt{1 - a^2})},
\end{equation} where \(0 \leq a \leq 1\) is the rotation parameter. Notice that upon putting \(a = 0\) we get \(T_{\rm H} = 1/(8 \pi M)\) and \(\Omega_{\rm H} = 0\), corresponding to the Schwarzschild black hole. Eq. (\ref{rotatingtempexp}) lets us write \(\Omega_{\rm H} = \frac{2 \pi T_{\rm H}a}{\sqrt{1 - a^2}}\), implying that
\begin{equation}
\Delta E = T_{\rm H} \bigg( \ln 2 + \frac{4 \pi a}{\sqrt{1 - a^2}}\bigg),
\end{equation} where we chose \(\alpha = 4\ln 2\) for which a transition between the consecutive area levels amounts to gain/loss of one bit of information. The energy loss exceeds Landauer's bound and is therefore consistent with Landauer's principle; the bound is saturated for \(a = 0\) which corresponds to the Schwarzschild black hole. It would be interesting to develop an interpretation of the loss of mass of rotating black holes as a result of superradiance along the lines of Landauer's principle; we keep this issue for future work. \\

\textbf{Acknowledgements:} B.B. thanks Brainware University for infrastructural support. A.G. is thankful to Chandrasekhar Bhamidipati, Sudipta Mukherji, and Narayan Banerjee for useful discussions and is supported by the Ministry of Education, Government of India in the form of a Prime Minister's Research Fellowship (ID: 1200454). S.S. is grateful to the Shiv Nadar Institution of Eminence for providing financial support and is also thankful to the Council of Scientific and Industrial Research (CSIR), Government of India [through Grant No. 09/1128(18274)/2024-EMR-I] for providing him with Direct-SRF fellowship.

\end{document}